\documentclass[aps,pra,onecolumn,showpacs,superscriptaddress,groupedaddress]{revtex4-1}
\usepackage[dvips]{graphics,graphicx}
\usepackage{epstopdf}
\usepackage{bbold}
\usepackage[usenames, dvipsnames]{color}
\usepackage{lipsum}
\usepackage{amssymb}
\usepackage{flushend}

\begin{document}
\title{Nonlinear response from optical bound states in the continuum}
\author{Evgeny N. Bulgakov$^{1,2}$}
\author{Dmitrii N. Maksimov$^{1,2}$}
\affiliation{$^1$Reshetnev Siberian State University of Science and Technology, 660037, Krasnoyarsk,
Russia\\
$^2$Kirensky Institute of Physics, Federal Research Center KSC SB
RAS, 660036, Krasnoyarsk, Russia}
\date{\today}
\begin{abstract}
We consider nonlinear effects in scattering of light by a periodic structure supporting optical bound states in
the continuum. In the spectral vicinity of the bound states the scattered electromagnetic field is resonantly
enhanced triggering optical bistability. Using coupled mode approach we derive a nonlinear equation for the amplitude
of the resonant mode associated with the bound state. We show that such an equation for the isolated resonance can be
easily solved yielding bistable solutions which are in quantitative agreement with the full-wave solutions of Maxwell's equations.
The coupled mode approach allowed us to cast the problem into the form of a driven nonlinear oscillator and
analyze the onset of bistability under variation of the incident wave. The results presented drastically simplify the analysis of nonlinear Maxwell's equations
and, thus, can be instrumental in engineering optical response via bound states in the continuum.
\end{abstract}
\maketitle

%\begin{abstract}
%We consider nonlinear effects in scattering of light by a periodic structure supporting optical bound states in
%the continuum. In the spectral vicinity of the bound states the scattered electromagnetic field is resonantly
%enhanced triggering optical bistability. Using coupled mode approach we derive a nonlinear equation for the amplitude
%of the resonant mode associated with the bound state. We show that such an equation for the isolated resonance can be
%easily solved yielding bistable solutions which are in quantitative agreement with the full-wave solutions of Maxwell's equations.
%The coupled mode approach allowed us to cast the the problem into the form of a driven nonlinear oscillator and
%analyze the onset of bistability under variation of the incident wave. The results presented drastically simplify the analysis nonlinear Maxwell's equations
%and, thus, can be instrumental in engineering optical response via bound states in the continuum.
%\end{abstract}

\section*{Introduction}

Optical bound states in the continuum (BICs) are peculiar localized eigenstates of Maxwell's equations embedded in the
continuous spectrum of scattering solutions \cite{Hsu16}. In the recent decade BICs have been theoretically predicted
\cite{Venakides03, Marinica08, Monticone14,Yang2014, Bulgakov2014, Gao16, Ni16, Rivera16,
Monticone17} and experimentally observed  \cite{Plotnik2011, Weimann13, Hsu13, Vicencio15, Sadrieva17, Xiao17} in various
dielectric set-ups with periodical permittivity. The BICs in photonic systems have already found important applications in
enhanced optical absorbtion \cite{Zhang15}, surface enhanced Raman spectroscopy \cite{Romano18}, lasing \cite{Kodigala17},
sensors \cite{Yanik11, Romano18a}, and filtering \cite{Foley14}.

Spectrally, the optical BICs are exceptional potions of leaky bands above the line of light where
the the quality factor ($Q$-factor) diverges to infinity \cite{Hsu16}. By themselves the BICs are localized solutions
decoupled from any external waves incident on the system. However, even the slightest off-set from the
BICs point in the momentum space transforms the BICs into high-$Q$ resonant modes with unlimited $Q$-factor
as far as the material losses in the supporting structure are neglected. In other words the BICs are spectrally
surrounded by strong resonances which can be excited from the far-field to arbitrary high amplitude by tuning the
angle of incidence of the incoming wave \cite{Yuan17}. The excitation of the strong resonances results in {\it critical field enhancement}
\cite{Yoon15,Mocella15a} with the near-field amplitude controlled by the frequency and the angle of incidence of the incoming monochromatic wave.

In this paper we investigate the role of the critical field enhancement in activation of nonlinear optical effects due to the cubic Kerr
nonlinearity. The earlier studies on the nonlinear effects were mostly concentrated on
the BICs supported by microcavities coupled to waveguide buried in the bulk photonic crystals, where
the nonlinear effects of symmetry breaking \cite{Bulgakov11} and channel dropping \cite{Bulgakov13} were demonstrated.
More recently the focus has been shift towards much simplier systems such as arrays of dielectric rods \cite{Yuan16,Yuan17}
and dielectric gratings \cite{Krasikov18}. So far, the problem was approached from two differing directions,
full-wave modelling \cite{Yuan16, Yuan17} that relies on exact numerical solution of Maxwell's equations,
and phenomenological coupled mode approach \cite{Krasikov18} that employs a set of equation in form of environment coupled
nonlinear oscillators. The former approach provides the solutions of the Maxwell's equations via time expensive numerical
simulations with no insight into the physical picture of the effect while the latter relies on a set of unknown parameters
whose numerical values have to be specified by fitting to exact numerical solutions. Here we bring the two approaches
together by deriving the coupled mode equation for the amplitude of the high-$Q$ resonant mode in the spectral vicinity
of the BIC. Thus, the problem is cast into the form of a single driven nonlinear oscillator. We show that
all parameters such {\it nonlinear} coupled mode theory (CMT) can be easily derived from the solution of the {\it linear}
scattering problem, and demonstrate the validity of our approach by comparing the CMT solutions against full-wave simulations
data.

\section*{Scattering theory}

\begin{figure}
\includegraphics[width=0.45\textwidth,trim={4cm 12.5cm 4cm 12cm},clip]{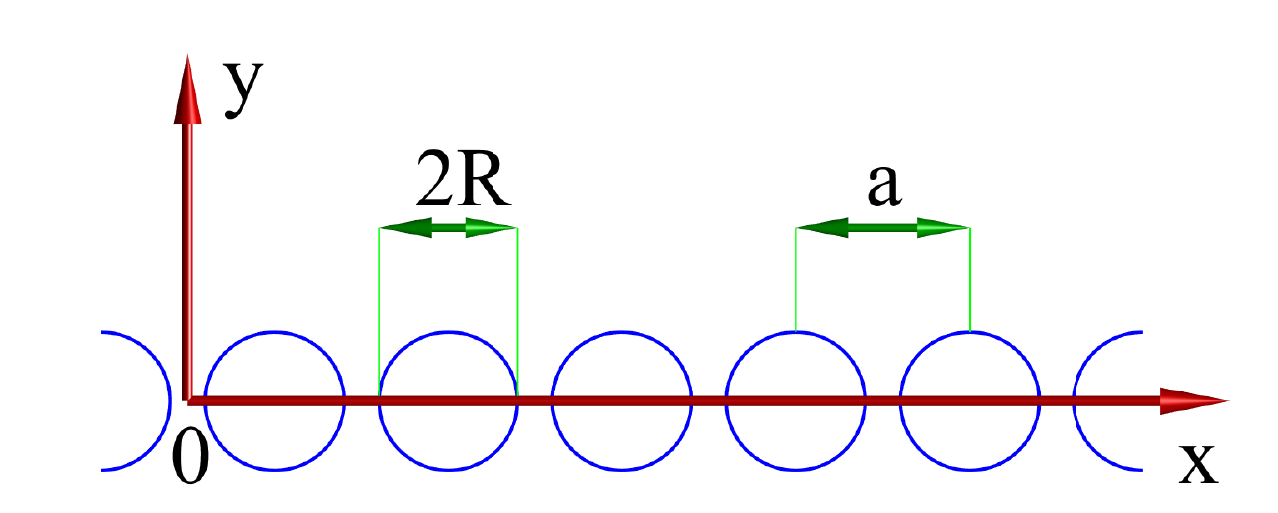}
\caption{Set-up of the array in $x0y$-plane. The circles show the surface cross-section of dielectric rods
with nonlinear permittivity.}
\label{fig1}
%%%%%%%%%%%%%%%%%%%%%%%%%%%%%%%%%%%%%%%%%%%%%%%%%%%%%%%%%%%%%%%%%%%%%%%%%%%%%%%%%%%%%%%%%%%%%%%%%%%%%%%%%%%
\end{figure}

We consider an array of identical dielectric rods of radius  R,  arranged along the x-axis with
period $a$. The axes of the rods are collinear and aligned with the $z$-axis. The cross-section of the
array in x0y -plane is shown in Fig. \ref{fig1}. The scattering problem is controlled by Maxwell's equation
which for the further convenience are written in the matrix form as follows
\begin{equation}\label{Maxwell}
\left\{
\begin{array} {cc}
0 & \nabla\times \\
-\nabla\times & 0
\end{array}
\right\}
\left\{
\begin{array} {c}
{\bf E}\\
{\bf H}
\end{array}
\right\}
=\frac{\partial}{\partial t}
\left\{
\begin{array} {c}
\epsilon{\bf E}\\
{\bf H}
\end{array}
\right\},
\end{equation}
where $\epsilon$ is the non-linear dielectric permittivity $\epsilon=n^2$ with $n$ as the refractive index
\begin{equation}\label{epsilon}
n=n_0+n_2I,
\end{equation}
where $n_0$ is the linear refractive index, $n_2$ is the nonlinear refractive index, and $I=|{\bf E}|^2$ is the intensity.
The scattering problem can be reduced to a single two-dimensional stationary differential equation
if monochromatic incident waves propagate in the directions orthogonal to the $z$-axis. In case
of $TM$-polarized waves that equation is written as
\begin{equation}\label{basic}
\frac{\partial^2 u}{\partial^2 x}+\frac{\partial^2 u}{\partial^2 y}+k_0^2\epsilon u=0,
\end{equation}
where $u$ is the $z$-component of the electric field $u=E_z$, and $k_0$ is the vacuum wave number (frequency).
Notice, that above we set the speed of light to unity to measure the frequency in the units of distance.
%%%%%%%%%%%%%%%%%%%%%%------------------FIG-1------------------------------%%%%%%%%%%%%%%%%%%%%%%%%%%%%%%%%
Assuming that a plane wave is incident from the upper half-space in Fig. \ref{fig1}.
the solution $y>R$ outside the scattering domain is written as
\begin{equation}\label{upper}
u(x,y)=\sqrt{2}\sum_{j=-\infty}^{\infty}r_je^{i[\alpha_j x+\beta_j(y-R)]}+\sqrt{2I_0}e^{i[\alpha_0\ x-\beta_0(y-R)]},
\end{equation}
where $\alpha_j=k_x+2\pi j/a$, $I_0$ is the intensity of the incident monochromatic wave, and $\beta_j=\sqrt{k_0^2-\alpha_j^2}$ with $k_x$ as the $x$-component of the incident
wave vector.
Tn the lower half-space we have
\begin{equation}\label{lower}
u(x,y)=\sqrt{2}\sum_{j=-\infty}^{\infty}t_je^{i[\alpha_j x-\beta_j(y+R)]}.
\end{equation}
The prefactor $\sqrt{2}$ in Eqs. (\ref{upper},\ref{lower}) is introduced to
have a unit period-averaged magnitude of the Poynting vector $\langle |{\bf S}| \rangle={\bf E}^{\dagger}{\bf E}/2=I_0$.

\begin{figure*}
\includegraphics[width=0.95\textwidth, height=0.53\textwidth,trim={1cm 9.5cm 1cm 9.5cm},clip]{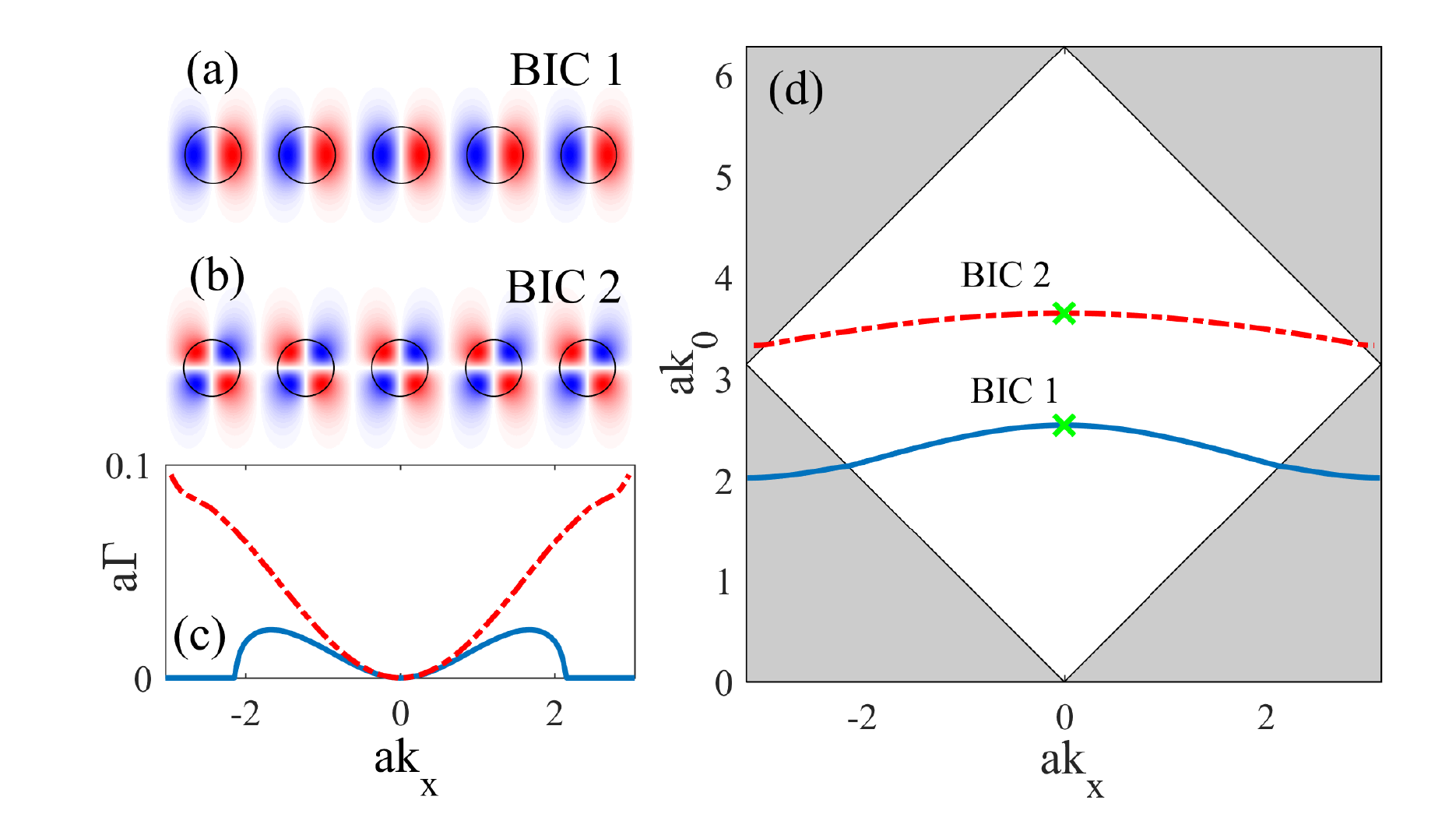}
\caption{BICs in the array of dielectric rods with $R=0.3a$, and dielectric permittivity within the
rods $\epsilon_1=12$. The ambient medium is air. (a,b) The
field profiles, $E_z(x,y)$ (p.d.u.) of BICs with eigenfrequencies $\bar{k}_{BIC}=2.5421a$, and $\bar{k}_{BIC}=3.6468a$. (c)
The dispersion of the imaginary part of the resonant eigenvalues.
(d) The real part of the resonant eigenvalue
of the leaky zones hosting the BICs; BIC 1 - solid blue, BIC 2 - dash red. The positions of the BICs are shown by green
crosses.}
\label{fig2}
\end{figure*}

The solution of the scattering problem is defined by the unknowns $t_j, r_j$ in Eqs. (\ref{upper},\ref{lower}).
Here for finding the BICs and the scattering solutions we applied a numerically efficient method based on the
Dirichlet-to-Neumann maps \cite{Huang06,Hu15}. We restrict ourselves with the simplest, namely, symmetry
protected BICs. Such BICs occur in the $\Gamma$-point as standing waves symmetrically mismatched
with outgoings waves with $k_x=0$. The field profiles of two such BICs
are shown in Fig. \ref{fig2} (a,b). The BICs are exceptional points of the leaky zones with a vanishing
imaginary part of the resonant eigenvalue $\bar{k}=\bar{k}_0-i\gamma$. The dispersions of the and real parts
of the resonant eigenvalue are shown in Fig. \ref{fig2} (c, d), respectively.

%%%%%%%%%%%%%%%%%%%%%%%----------------FIG-2-----------------%%%%%%%%%%%%%%%%%%%%%%%%%%%%%%%%%%%%%%%%%%%%%%%%%%%
%%%%%%%%%%%%%%%%%%%%%%%%%%%%%%%%%%%%%%%%%%%%%%%%%%%%%%%%%%%%%%%%%%%%%%%%%%%%%%%%%%%%%%%%%%%%%%%%%%%%%%%%%%%%%%%%%%

\begin{figure}
\includegraphics[width=0.93\textwidth,trim={0cm 8.5cm 0cm 9.cm},clip]{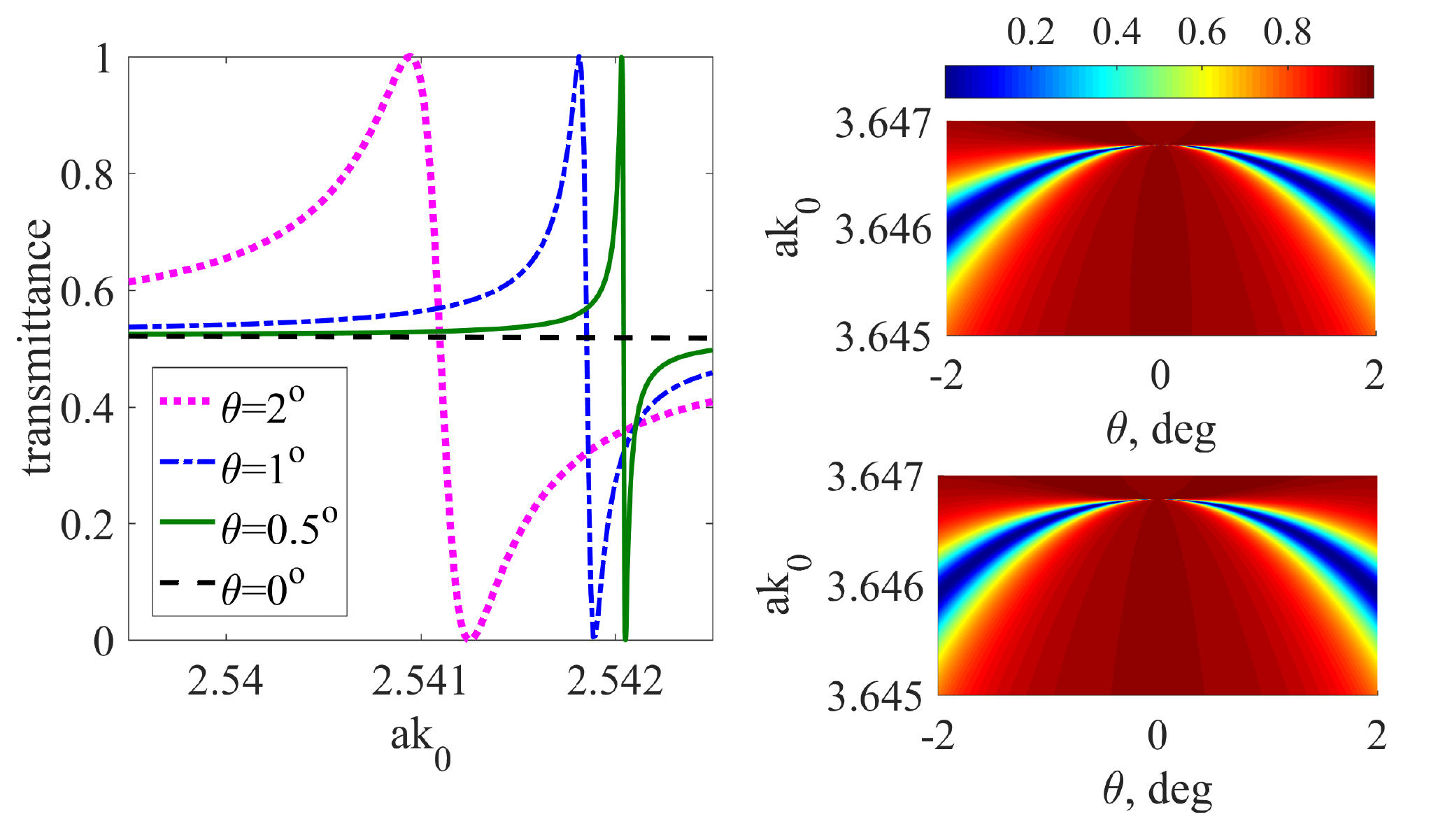}
\caption{Scattering of a monochromatic plane wave in the spectral vicinity of BICs, $n_0=12$ (polycrystalline silicon at 1.8 ${\rm \mu m}$
\cite{Li80}). (Left panel)
Collapsing Fano feature in transmittance in the spectral vicinity of BIC 1 from Fig. \ref{fig2}. (Right panel) Transmittance in the spectral vicinity of BIC 2;
top - full-wave solution, bottom - CMT approximation.}
\label{fig3}
\end{figure}

One important hallmark of the BICs is a narrow Fano feature in the transmittance spectrum with occurs in the spectral
vicinity of the BIC \cite{Kim,Shipman,SBR,Blanchard16} as the angle of incidence, $\theta=\arcsin(k_x/k_0)$ is slightly detuned from the normal. This effect is illustrated
in Fig. \ref{fig3} (left panel). One can see from Fig. \ref{fig3} that the presence of a BIC induces a Fano resonance that collapses
on approach to the normal incidence.

%%%%%%%%%%%%%%%%%%%%-------------------------FIG-3---------%%%%%%%%%%%%%%%%%%%%%%%%%%%%%%%%%%%%%%%%%%%%%%%%%%%%%%%
%%%%%%%%%%%%%%%%%%%%%%%%%%%%%%%%%%%%%%%%%%%%%%%%%%%%%%%%%%%%%%%%%%%%%%%%%%%%%%%%%%%%%%%%%%%%%%%%%%%%%%%%%%%%%%%%%%

To quantitatively describe the scattering in the spectral vicinity of the BICs we resort to coupled mode theory (CMT)
for a single isolated resonance \cite{Suh04}. According to CMT the amplitude of a leaky mode $c(t)$ obeys the
following temporal equation
\begin{equation}\label{CMT_linear}
\frac{d c(t)}{d t}=-(i\bar{k}_0+\gamma)c(t)+\varkappa \sqrt{I_0}e^{-ik_0t},
\end{equation}
where $\bar{k}_0, \gamma$ are given by dispersion relationships shown in Fig. \ref{fig2} (c, d), and
$\varkappa$ is the coupling coefficient. In the case of stationary scattering, $c(t)=ce^{-ik_0t}$ the scattering matrix is written as
\cite{Suh04}
\begin{equation}\label{S_matrix}
\widehat{S}=\widehat{C}+\frac{{\bf d}{\bf d}^T}{i(\bar{k}_0-k_0)+\gamma},
\end{equation}
where $\widehat{C}$ is the matrix of the direct process, and ${\bf d}^T=[\varkappa, \pm \varkappa]$,
the sign $+(-)$ being chosen if the mode is symmetric (antisymmetric)
with respect $y \rightarrow -y$, see Fig. \ref{fig2} (a, b).
By applying energy conservation it can be shown \cite{Suh04} that ${\bf d}^{\dagger}{\bf d}=2\gamma$,
therefore $\varkappa=e^{i\delta}\sqrt{\gamma}$. In addition the time reversal yields $\widehat{C}{\bf d}^{*}=-{\bf d}$.
Since $\widehat{C}$ is symmetric the latter constraint uniquely defines the phase $\delta$. The spectrum in Figs. \ref{fig2} (c, d)
is symmetric with respect to $k_x \rightarrow -k_x$, hence in the vicinity of the symmetry protected BICs we can write \cite{Bulgakov17oe}
\begin{eqnarray}\label{expansion}
a\bar{k}_0(\theta)=a\bar{k}_{BIC}+a_2\theta^2+a_4\theta^4+{\cal O}(\theta^6), \nonumber \\
a\gamma(\theta)=b_2\theta^2+b_4\theta^4+{\cal O}(\theta^6).
\end{eqnarray}
In Table \ref{Table} we collect the values of all parameters necessary for finding transmittance and reflectance with equation (\ref{S_matrix}).
The parameters $a_2, a_4, b_2, b_4$ are extracted by the  least square fit in the vicinity of the BIC, while the entries of $\widehat{C}$ are found
at the normal incidence and the BIC frequency of the incident wave. In Fig. \ref{fig3} (right panel) we plot the transmittance in the spectral
vicinity of BIC 2 obtain thorough full-wave modelling in comparison against the CMT fit. One can see that the CMT reproduces the full-wave solution to a good
accuracy.
 %\begin{table}
%\begin{center}
%\begin{tabular}{lcc}
%\ & BIC 1 & BIC 2  \\
%\hline
%$a\bar{k}_{BIC}$ & 2.542108  & 3.646776  \\
%$\{\widehat{C}\}_{1,1}$ & \ $-0.421696-0.551370i$  \   & \ $0.1568749+0.0586432i$  \      \\
%$\{\widehat{C}\}_{1,2}$ & \ $0.571777-0.437304i$   \   & \ $-0.345210+0.923462i$   \     \\
%$a_2$ & $-2.5403\times10^{-4}$     & $-1.6338\times10^{-4}$     \\
%$a_4$ & $1.3313\times10^{-6}$     & $-8.7802\times10^{-6}$         \\
%$b_2$ & $3.7506\times10^{-5}$    &  $8.9939\times10^{-5}$       \\
%$b_4$ & $-2.7612\times10^{-8}$     & $-3.1609\times10^{-8}$       \\
%\end{tabular}
%\caption{Parameters of the scattering theory in the spectral vicinity of BIC 1 and BIC 2.}
%\label{Table}
%\end{center}
%\end{table}

\begin{table}
\begin{center}
\begin{tabular}{cccccccc}
BIC & $a\bar{k}_{BIC}$ & $\{\widehat{C}\}_{1,1}$ & $\{\widehat{C}\}_{1,2}$ & $a_2\times10^{4}$ & $a_4\times10^{6}$ & $b_2\times10^{5}$ & $b_4\times10^{8}$ \\
\hline
1 & 2.54211  & $-0.42170-0.55137i$   & $0.57178-0.43730i$   & $-2.5403$ & $1.3313$  & $3.7506$ & $-2.7612$ \\
2 & 3.64678  &  $0.15687+0.05864i$   &  $-0.34521+0.92346i$  & $-1.6338$ & $-8.7802$ & $8.9939$ & $-3.1609$
\end{tabular}
\caption{Parameters of the scattering theory in the spectral vicinity of BIC 1 and BIC 2.}
\label{Table}
\end{center}
\end{table}

\section*{Effect of the nonlinearity}
The effect of the nonlinearity can be incorporated to the time-stationary CMT equation by
introducing nonlinear frequency shift $\Delta k_0$ due to the Kerr effect
\begin{equation}\label{CMT_nonlinear_stationary}
[i(\bar{k}_0-\Delta k_0-k_0)+\gamma] c=\varkappa \sqrt{I_0},
\end{equation}
where $\Delta k_0$ is dependent on $c$. The perturbative frequency shift induced by variation of dielectric constant
can be found as \cite{Soljacic02, Koenderink05, Joannopoulos11, Ramunno09}
\begin{equation}
\Delta k_0=\frac{\lambda}{4}|c|^2,
\end{equation}
where
\begin{equation}\label{lambda}
\lambda=2n_0n_2\int\limits_{S_{R}} dS |{\bf E}_{BIC}|^4,
\end{equation}
with integration performed over the cross section of the dielectric rod, $S_R$ and the BIC field ${\bf E}_{BIC}$ normalized
to store a unit period averaged energy
\begin{equation}\label{normalization}
\int\limits_S dS \frac{n_0(x,y)^2{\bf E}^{\dagger}{\bf E}+{\bf H}^{\dagger}{\bf H}}{4}=1,
\end{equation}
where $S$ is the area of the elementary cell. Although equation (\ref{lambda}) is known to have certain limitations for low-$Q$ cavities \cite{Lalanne18}, it is found to be applicable for high-$Q$ nonlinear
cavities embedded into the bulk photonic crystals \cite{Soljacic02}.

In more detail, to introduce the effect of nonlinearity into CMT we decompose the electromagnetic field into two components
${\bf E}={\bf E}_{res}+{\bf E}_{dir}, {\bf H}={\bf H}_{res}+{\bf H}_{dir}$. Here subscript
$dir$ designates the direct field contribution associated with the non-resonant optical pathway through the structure,
while subscript $res$ is used for the contribution due to resonant excitation of the leaky wave which evolves to a BIC
at the normal incidence, see Fig. \ref{fig2} (c, d). Substituting the decomposed field into Maxwell's equations, equation (\ref{Maxwell}) one finds
%\begin{widetext}
\begin{equation}\label{bigeqution}
\left\{
\begin{array} {cc}
0 & \nabla\times \\
-\nabla\times & 0
\end{array}
\right\}
\left\{
\begin{array} {c}
{\bf E}_{res}\\
{\bf H}_{res}
\end{array}
\right\}
-\frac{\partial}{\partial t}
\left\{
\begin{array} {c}
\epsilon{\bf E}_{res}\\
{\bf H}_{res}
\end{array}
\right\}=
\left\{
\begin{array} {c}
-\nabla\times{\bf H}_{dir}\\
\nabla\times{\bf E}_{dir}
\end{array}
\right\}
+\frac{\partial}{\partial t}
\left\{
\begin{array} {c}
\epsilon{\bf E}_{dir}\\
{\bf H}_{dir}
\end{array}
\right\}.
\end{equation}
%\end{widetext}
\begin{figure*}
\includegraphics[width=0.95\textwidth, height=0.83\textwidth,trim={2cm 7cm 2cm 7cm},clip]{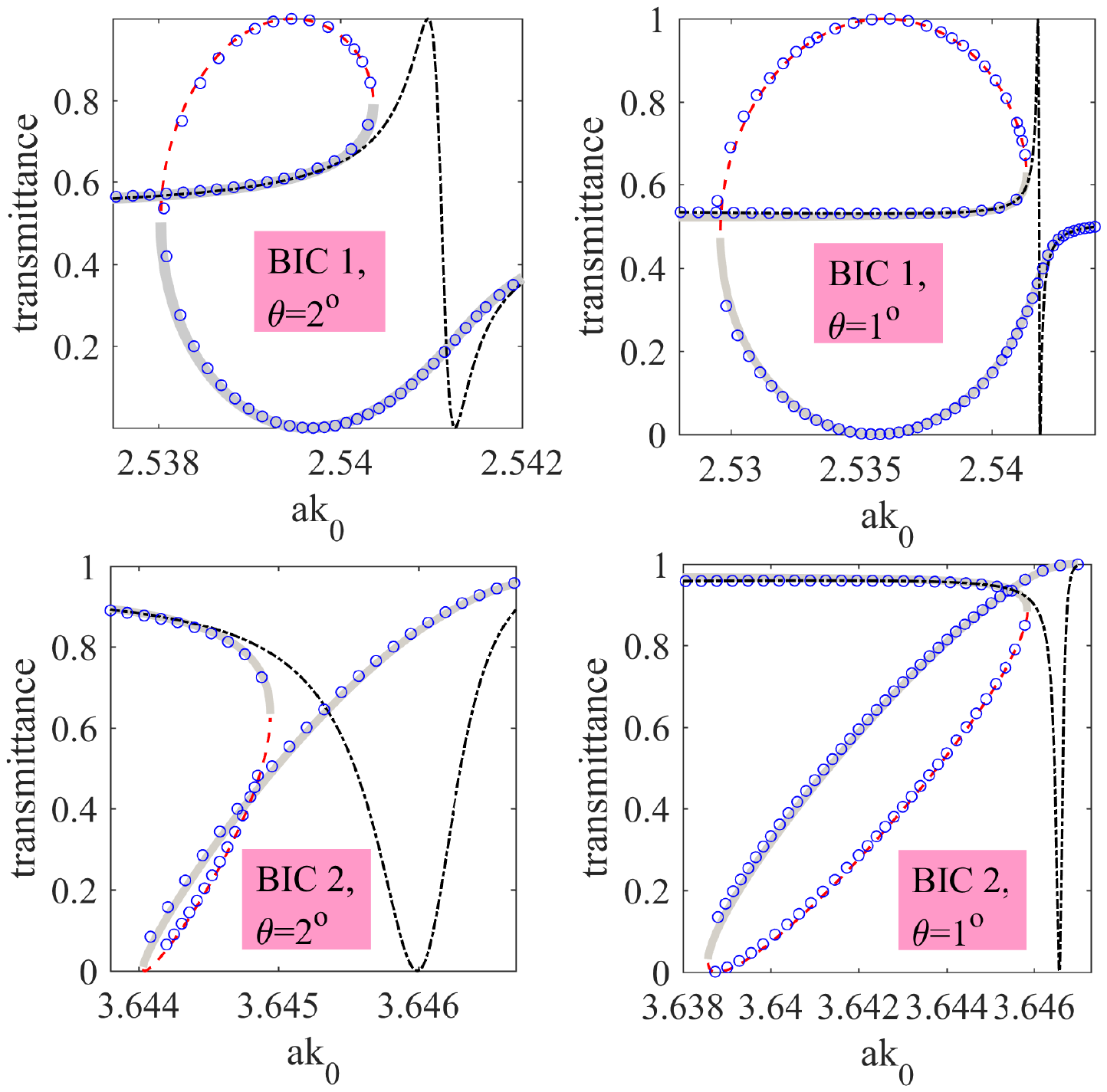}
\caption{Nonlinear Fano resonance in the spectral vicinity of BIC 1 and BIC 2 at different angles of incidence, $\theta$ for $I_0=8.(3) {\rm MW/cm^2}$. Blue circles - numerical
results by Fourier-Chebyshev pseudospectral method, thick gray line - stable CMT solution, thin dash red line - unstable CMT solution, dash-dot black
line - Fano line-shape unperturbed by the nonlinearity.}
\label{fig4}
\end{figure*}
The temporal dependance of the resonant contribution can be written as ${\bf E}_{res}(t)=c(t){\bf E}_{0}$,
${\bf H}_{res}(t)=c(t){\bf H}_{0}$, where ${\bf E}_0, {\bf H}_0$ are the electric and magnetic field profiles of the leaky mode.
Multiplying from the left by $1/4 [{\bf E}_{0}^{\dagger}, {\bf H}_{0}^{\dagger}]$ and integrating over the scattering domain
one immediately finds
\begin{equation}\label{CMT_non-linear}
\frac{d }{d t}\left(c(t)+\frac{\lambda}{4}|c(t)|^2c(t) \right)=-(i\bar{k}_0+\gamma)c(t)+be^{-ik_0t}, \
\end{equation}
where we assumed that ${\bf E}_{dir}, {\bf H}_{dir}$ are monochromatic fields with frequency $k_0$, and neglected the nonlinear
effects in the direct field since its amplitude is much smaller than that of the resonant field. We also assumed
that the leaky mode is normalized according to equation (\ref{normalization})
to be consistent we our normalization of the outgoing waves Eqs. (\ref{upper}, \ref{lower}).
By comparing equation (\ref{CMT_non-linear})
against equation (\ref{CMT_linear})
we find
\begin{equation}
b=\varkappa \sqrt{I_0}.
\end{equation}
 The only problem we left with is to correctly define $\lambda$. We have mentioned
that equation (\ref{CMT_non-linear}) is obtained after integration over the scattering domain which is somewhat ambiguous
since the boundary between the far- and near-fields can arbitrary defined. What is worst is that the resonant eigenmodes diverge in the far-field,
and therefore require a different normalization condition \cite{Doost14} rather than equation (\ref{normalization}). One may notice,
however, that evaluation of $\lambda$ in equation (\ref{CMT_non-linear}) can only involve integration over the area of the rods where
the non-linearity is present. One the other hand the leaky mode is spectrally close to the BIC, hence we conjecture that the leaky
mode field profile within the rods can be replaced with that of the BIC. This approach lifts the problem of the mode normalization
as the BIC is a localized state square integrable over the whole space. Thus, we end up with equation (\ref{lambda}).

\begin{figure*}
\includegraphics[width=0.9\textwidth, height=0.45\textwidth,trim={0cm 9.5cm 1cm 10cm},clip]{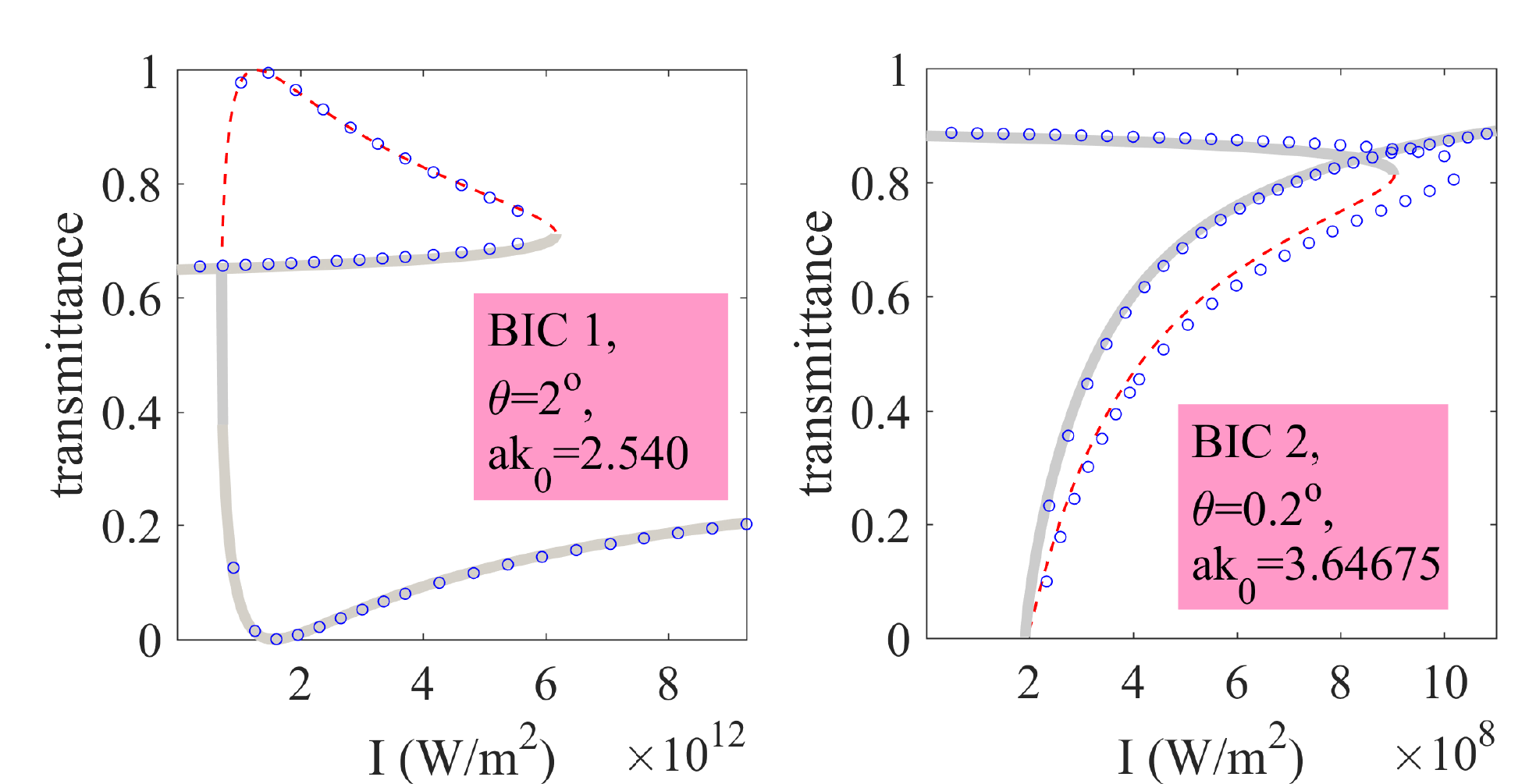}
\caption{Optical bistability in the intensity domain with BIC 1 and BIC 2. Blue circles - numerical
results by Fourier-Chebyshev pseudospectral method, thick gray line - stable CMT solution, thin dash red line - unstable CMT solution.}
\label{fig5}
\end{figure*}
After time harmonic substitution, $c(t)=ce^{-ik_0t}$, equation (\ref{CMT_non-linear}) can be solved for the system's response to a monochromatic wave.
The transmission amplitude can be found as \cite{Suh04}
\begin{equation}
t_0=\{\widehat{C} \}_{1,2}\sqrt{I_0}+\varkappa c.
\end{equation}
The stability of time harmonic solutions can be examined with a small perturbation approach which yields that the solution is
stable if and only if the real part of both eigenvalues of the matrix
%\begin{widetext}
\begin{equation}\label{stability}
\widehat{M}=%\frac{1}{1+\lambda|a|^2+\frac{3\lambda^2|a|^4}{16}}
\left\{
\begin{array} {cc}
1+\frac{\lambda|c|^2}{2} & - \frac{\lambda c^2}{4}\\
- \frac{\lambda (c^{*})^2}{4} & 1+\frac{\lambda|c|^2}{2}
\end{array}
\right\}
\left\{
\begin{array} {cc}
ik_0\frac{\lambda|c|^2}{2}-i(\bar{k}_0-k_0)-\gamma & ik_0 \frac{\lambda c^2}{4}\\
-ik_0 \frac{\lambda (c^{*})^2}{4} & -ik_0\frac{\lambda|c|^2}{2}+i(\bar{k}_0-k_0)-\gamma
\end{array}
\right\}
\end{equation}
%\end{widetext}
are non-positive.

Finally, we verified our findings by comparing the solution of equation (\ref{CMT_non-linear}) against exact numerical solutions of equation (\ref{basic}) obtained
with Fourier-Chebyshev pseudospectral method \cite{Yuan13}. For our numerical simulations we took $n_2=5\times10^{-18} {\rm m^2/W}$ which corresponds
to silicon at $1.8 {\rm \mu m}$ \cite{Yue12}. The results are shown in Fig. \ref{fig4} where one can see a good agreement between the two approaches. In
Fig. \ref{fig2} one can see the typical picture of nonlinear Fano resonances \cite{Miroshnichenko05} with optical bistability triggered by critical field
enhancement in the spectral vicinity of a BIC \cite{Yoon15,Mocella15a}. Notice that the stability pattern is identical to that previously reported
in the literature \cite{Miroshnichenko05, Krasikov18}. We also investigated the emergence of optical bistability in the intensity. The simulations
were again performed by both solving equation (\ref{CMT_non-linear}), and solving equation (\ref{basic}) by full-wave Fourier-Chebyshev pseudospectral method. In Fig.
\ref{fig5} (left panel)
we show a picture of optical bistability in the spectral vicinity of BIC 1. Notice, that the bistability widow occurs at the intensities unobtainable with
with $1W$ continuous lasers. To reduce the bistability threshold one can tune the angle of incidence approaching the BIC in the momentum space and, thus,
increasing the $Q$-factor of the leaky mode \cite{Yuan17}. This idea is exemplified in Fig. \ref{fig5} (right panel) where we plot the transmittance in
the spectral vicinity of BIC 2 at the incident angle $\theta=0.2 \ {\rm deg}$. One can see that the window of optical bistability is now
$(0.3-1.5) \times 10^4 \ {\rm W/cm^2}$.

The bistability threshold can be accessed by equating the resonant width $\gamma$ to the frequency shift induced by the nonlinearity $\Delta k_0$ at the
spectral point of maximal resonant enhancement. That yields  $a\gamma=(\lambda/4)I_0/(a\gamma)$. By applying equation (\ref{expansion}) up to the term quadratic in $\theta$ one
finds
\begin{equation}\label{threshold}
I_0=\frac{4}{\lambda}b_2^2\theta^4.
\end{equation}
One can see from equation (\ref{threshold}) that as far as the material losses are neglected there is no intensity threshold for optical bistability
induced by BICs.
This result is, however, achieved at the cost of a precise control of the frequency of the incident wave so that the line width of
the continuous laser  has to smaller than the resonant width $\gamma$,
hence we have seven significant digits in the inset in Fig. \ref{fig5} (right panel).
Theoretically, any arbitrary low threshold of optical bistability can be achieved by decreasing the angle of incidence once material losses,
thermooptical effects and structure fabrication inaccuracies are neglected. In a realistic physical experiment, though, engineering optical
set-ups for observing bistability with a BIC will always be a trade-off between the line width and the intensity of the laser available, as well as, should
take into account thermal deformation of the structure due to heating and fabrication inaccuracies limitations on the $Q$-factor.

\section*{Acknowledgements}

This work was supported by Ministry of Education and Science of Russian Federation
(state contract N 3.1845.2017/4.6). We appreciate discussions with Ya Yan Lu, Lijun Yuan,
Andrey M. Vyunishev, and Ivan V. Timofeev.

\section*{Discussion}

We have theoretically shown the effect optical bistablity with bound states in the continuum (BIC). The physical picture of the effect is explained
through coupled mode theory which allowed us to cast the problem of optical response to the simple form of a single driven nonlinear
oscillator. The proposed coupled mode approach reduces the problem to finding the solution of the {\it linear} Maxwell's equation
in the spectral vicinity of the BIC. Then, all parameters entering the {\it nonlinear} coupled mode equation can be easily found
from the dispersion of the leaky band hosting the BIC, the scattering matrix of the direct process, and the BIC mode profile.
The proposed method enormously simplifies analyzing the nonlinear effects induced by bound states in the continuum since it makes
possible to avoid time expensive full-wave simulations. The resulting picture of a nonlinear Fano resonance can be easily understood in terms
of a frequency shift due to the Kerr nonlinearity activated by critical field enhancement in the spectral vicinity of a BIC. We believe
that the results will be of use in engineering optical set-ups for observation nonlinear effects with BICs.

\bibliography{BSC_light_trapping}

%\noindent LaTeX formats citations and references automatically using the bibliography records %in your .bib file, which you can edit via the project menu. Use the cite command for an inline %citation, e.g.  \cite{Hao:gidmaps:2014}.

%For data citations of datasets uploaded to e.g. \emph{figshare}, please use the %\verb|howpublished| option in the bib entry to specify the platform and the %link, as in the \verb|Hao:gidmaps:2014| example in the sample bibliography %file.

\end{document}